\documentstyle[aps,epsfig,prd,floats,preprint]{revtex}

\preprint{ UCRHEP-T232\quad BNL-HET-98/26}

\def\gsim{\lower0.5ex\hbox{$\:\buildrel >\over\sim\:$}}
\def\lsim{\lower0.5ex\hbox{$\:\buildrel <\over\sim\:$}}

\def \sneum{{\tilde\nu}^\mu}
\def \sneut{{\tilde\nu}^\tau}

\def \sneump{{\tilde\nu}^\mu_+}
\def \sneumm{{\tilde\nu}^\mu_-}
\def \sneuip{{\tilde\nu}^i_+}
\def \sneuim{{\tilde\nu}^i_-}
\def \sneums{{\tilde\nu}^\mu_s}

\def \sneui{{\tilde\nu}^i}

\def \sneupm{{\tilde\nu}_{\pm}}
\def \sneutpm{{\tilde\nu}^\tau_{\pm}}
\def \sneuipm{{\tilde\nu}^i_{\pm}}
\def \sneumpm{{\tilde\nu}^\mu_{\pm}}
\def \rp{{R\hspace{-0.22cm}/}_P}
\def \lp{{L\!\!\!/}}

\def \be {\begin{equation}}
\def \ee {\end{equation}}
\def \bea{\begin{eqnarray}}
\def \eea{\end{eqnarray}}

\def \n{\noindent}

\begin{document}

\title{\boldmath{$R$}--parity violation and uses of the 
rare decay
\boldmath{$\tilde\nu \to \gamma \gamma$} in hadron and photon colliders}

\author{
S. Bar-Shalom, G. Eilam\thanks{On leave from: Physics Department, Technion-Institute 
of Technology, Haifa 32000,  Israel.}, 
J. Wudka}
\address{
Physics Department, University of California, Riverside, 
CA 92521, USA}
\bigskip

\author{A. Soni}
\address{Physics Department, Brookhaven National Laboratory, Upton, NY 11973,
USA}
\vspace{.2in}
\date{\today}

\maketitle

\begin{abstract}

We consider implications of the loop process 
$\tilde\nu \to \gamma \gamma$  
in the MSSM with $R$-parity violation ($\rp$) for future experiments, 
where the 
sneutrino is produced as the only
supersymmetric particle.
We present a scenario 
for the $\rp$ couplings, where
this clean decay, although rare with 
${\rm Br}(\tilde\nu \to \gamma \gamma) \sim 10^{-6}$, 
may be useful for sneutrino detection
over a range of sneutrino masses
at the LHC. 
Furthermore,
the new $\tilde\nu \gamma \gamma$ effective coupling   
may induce detectable sneutrino resonant production in $\gamma \gamma$
collisions, over a considerably wide mass range. We compare 
$\tilde\nu \to \gamma \gamma,~ gg$ throughout the paper 
with the analogous yet quantitatively
very different, Higgs $\to \gamma \gamma,~gg$ decays and 
comment on the loop processes $\tilde\nu \to WW,~ZZ$.

\end{abstract}

\newpage

\n The generalization of the MSSM which includes
$R$-parity violating ($\rp$) processes has been gaining increasing 
attention in the past few years \cite{rpreview}. 
The presence of $\rp$ couplings 
drastically changes the 
phenomenology of supersymmetric theories, by opening 
new experimental strategies in the search for supersymmetry.
The sneutrino sector of the MSSM, in which we are interested 
here, can exhibit new phenomena directly related to the 
lepton violating $\rp$ operators, {\it e.g.,} sneutrinos can be 
produced as $s$--channel resonances 
\cite{resold,sneuresonant,ourpapers}, sneutrinos and anti--sneutrinos can mix 
\cite{sneumixing} and the sneutrino mixing phenomenon can drive 
large tree--level CP--violating asymmetries \cite{ourpapers}. 

In this paper we study another issue in sneutrino physics 
unique to the MSSM with $\rp$, namely the role of rare sneutrino decays in 
collider experiments. As is well known 
rare decays can play a crucial role in 
collider experiments. An example is the rare Higgs decay \cite{foot1} 
$h \to \gamma \gamma$, 
which has a branching ratio of ${\cal O}(10^{-3})$ for $m_h \lsim 2m_W$ 
\cite{hepph9803257}. 
In spite of this small branching ratio, it is now 
widely believed that 
this rare decay mode may be the best discovery channel for Higgs with a 
mass $\lsim 140$ GeV at the LHC. It also has implications for Higgs
production in $\gamma \gamma$ collisions. On the other hand,
the effective $hgg$ coupling ($g$=gluon) is unimportant for 
discovery of $h$
in view of the large QCD background, but is 
believed to be 
the main mechanism  for  Higgs 
production at the LHC.

Here we will concentrate on the decay 
$\sneui \to \gamma \gamma$, where 
$i=e,\mu,\tau$ indicates the sneutrino flavor, 
and briefly comment on the other 
rare decay channels $\sneui \to gg,~ZZ,~W^+W^-$.
These $\rp$ sneutrino decays into vector bosons, occur at the one loop--level 
with an insertion of one $\rp$ sneutrino coupling to down--quarks or leptons. 
The decay of a sneutrino to a pair of photons, $\tilde\nu \to \gamma \gamma$, 
in the MSSM with 
$\rp$, while resembling 
$h \to \gamma \gamma$, has its own unique characteristics. 
In fact, as will be shown in this paper, 
although the branching ratio of $\tilde\nu \to \gamma \gamma$ is much 
smaller than that of $h \to \gamma \gamma$, it may be compensated 
by a large sneutrino production rate as compared to the Higgs 
case at the LHC. The basic reaction that we will consider is the inclusive, 
single $\tilde\nu$ production $pp \to \tilde\nu +X$ via the parton processes
$b \bar b \to \tilde\nu$, $b\;({\rm or}~\bar b)\,g \to \tilde\nu + 
b\;({\rm or}~\bar b)\,$, 
$b \bar b \to \tilde\nu+g$ and $gg \to \tilde\nu$, all followed by 
$\tilde\nu \to \gamma \gamma$.
At the LHC this $\gamma \gamma$ mode is found to be 
useful as a sneutrino discovery channel over a sneutrino mass range
approximately equal to
the corresponding Higgs mass range ({\it i.e.}, $m_h \lsim 140$ GeV). 
In addition, both $h$ and 
$\tilde\nu$ can be produced in $\gamma \gamma$ collisions, though with a 
smaller rate for $\tilde\nu$.

The relevant $\rp$ Lagrangian is \cite{rpreview}:

\be
{\cal L}_{\lp} = \frac{1}{2}\lambda_{ijk} {\hat L}_i  
{\hat L}_j {\hat E}_k^c + \lambda_{ijk}^{\prime} {\hat L}_i  
{\hat Q}_j {\hat D}_k^c
\label{rparity}~,     
\ee 

\n where ${\hat L}$ and ${\hat Q}$ are the SU(2)--doublet lepton and quark 
superfields, respectively and ${\hat E}^c$ and ${\hat D}^c$ are the 
lepton and quark singlet 
superfields, respectively. Also, the flavor indices $i,j,k$ are such that, 
for the pure leptonic operator in Eq.~\ref{rparity},  
$i \neq j$. Throughout this paper we will neglect another possible 
lepton violating $\rp$ term of the
form $ L_i H_u $ in the superpotential; the effects of such a term have been
considered elsewhere (see \cite{rpreview} and references therein).

In order to calculate the decay rate of $\sneui \to \gamma \gamma$ it is 
convenient to define $\sneui = (\sneui_+ + i \sneui_-)/\sqrt 2$ 
and work in 
the $\sneuipm$ mass basis. The relevant $\rp$ 
couplings of $\sneuipm$ to down--quarks and leptons are then given by:

\bea 
&& \sneuip d_j d_k :\ \  i \lambda^{\prime}_{ijk}/\sqrt 2~~,~~
\sneuim d_j d_k :\ \  - \lambda^{\prime}_{ijk} \gamma_5 /\sqrt 2~,\\
&& \sneuip \ell_j \ell_k :\ \  i \lambda_{ijk}/\sqrt 2~~,~~
\sneuim \ell_j \ell_k :\ \  - \lambda_{ijk} \gamma_5 /\sqrt 2~,
\eea

The calculation can now be simply performed in analogy with the CP--even 
($h$) and
CP--odd ($A$) neutral Higgs decays to a pair of photons in the MSSM 
\cite{higgshunters} (and similarly for a pair of gluons):

\bea
\Gamma(\sneuipm \to \gamma \gamma) &=& \frac{\alpha^2 m^3_{\sneuipm}}
{512 \pi^3} \sum_{j=1}^{3} \left|\frac{N_c}{m_{d_j}} e_{d_j}^2
\lambda^{\prime}_{ijj} F_{1/2}^\pm (\tau_{d_j}) \right. \nonumber \\
&+& \left. \frac{1}{m_{\ell_j}} \lambda_{ijj} F_{1/2}^\pm (\tau_{\ell_j}) 
(1-\delta_{ij}) \right|^2 
\label{gammagamma}~,
\eea

\bea
\Gamma(\sneuipm \to gg) &=& \frac{\alpha_s^2 m^3_{\sneuipm}}
{256 \pi^3} \sum_{j=1}^{3} \left|\frac{1}{m_{d_j}}\lambda^{\prime}_{ijj} 
F_{1/2}^\pm (\tau_{d_j}) \right|^2 
\label{gluongluon}~,
\eea

\n where in Eq.~\ref{gammagamma} the sum runs over all down--fermions,
while in Eq.~\ref{gluongluon} only down--quarks are included.
Furthermore, $N_c=3$ is the number of colors, $e_{d_j}=-1/3$
is the charge of quark $d_j$. The functions $F_{1/2}^\pm(\tau)$, 
where $\tau=4 m^2/m^2_{\tilde{\nu}}$ , are defined as follows 
\cite{foot2}:

\be
F_{1/2}^+ = -2\tau\left[1+(1-\tau) f(\tau)\right]~, \qquad
F_{1/2}^- = -2\tau f(\tau) \label{fhalf} ~,
\ee

\n where:

\be
f(\tau)= \left\{ \begin{array}{ll}
\left[\sin ^{-1}\left(\sqrt{1/\tau}\right)\right]^2~, & \mbox{if $\tau\geq 1$}
~,\vspace{1em}\\
     -\frac{1}{4}\left[\ln\left(\eta_+/\eta_-\right)-i\pi\right]^2~,
& \mbox{if $\tau < 1$}
\end{array}\right. 
\ee

\n and:

\be
\eta_\pm \equiv 1 \pm \sqrt{1-\tau}~. 
\ee

\n Note that $F_{1/2}^\pm (m)/m \to 0$ for $m \to 0$.

 From the above equations we observe that, unlike the Higgs case,
$W$ bosons, charged Higgs particles (present in some extensions of the SM),
2/3 charged quarks and neutrinos 
do not appear in the loop.
This results from the absence of the relevant terms in the     
$\rp$ Lagrangian. 
In addition, sfermions are excluded from the loop.
This is similar to $A \to \gamma \gamma,~gg$ but, unlike 
the corresponding decays of $h$ in the MSSM \cite{higgshunters},
is due to the chirality conserving $\gamma$ (and $g$)
coupling to sfermions \cite{haberkane}.
 
We now describe two possible scenarios,
each one of which 
has distinct phenomenological implications
for collider experiments.

\begin{itemize}

\item \underline{Scenario 1: Only $\lambda^{\prime}_{i33} \neq 0$ and
$\lambda_{i33} \neq 0$}

Within this scenario, which may be theoretically motivated by imposing a mass
hierarchy on the $\rp$ couplings in (\ref{rparity}) ({\it i.e.,} only 
the heavier third generation fermions have a non--negligible $\rp$ 
coupling to sneutrinos), the sneutrinos 
can be produced as the single supersymmetric particle at 
the LHC and in a future $\gamma \gamma$ linear collider
(Photon Linear Collider, or PLC \cite{foot3}),
but not in $e^+e^-$ and $\mu^+ \mu^-$ colliders.

As will be shown below, 
due to the high production rate of $\sneuipm$, predominantly  
through the $b \bar b$ and $bg$ fusion processes, 
the decay $\sneuipm \to \gamma \gamma$ may prove to be a useful 
detection mechanism ({\it \`{a} la} $h \to \gamma \gamma$) 
at the LHC. 
At the Tevatron, due to the low $b$--quark and gluon content of the beams,
the $\gamma \gamma$ decays of sneutrinos cannot be used to
detect them; as we comment later, it appears difficult 
to ``save'' this detection mode 
at the Tevatron even in 
a modified scenario.

\item \underline{Scenario 2: All $\lambda^{\prime}_{ijk}=0$
and only $\lambda_{i33} \neq 0$}

In this scenario the only $ \rp $ interactions present couple
$\sneui$ with $i=e,\mu$ to a pair of $\tau$ leptons.
In this case sneutrinos will not be produced singly 
in either $e^+e^-$, $\mu^+\mu^-$ colliders or 
in hadron colliders \cite{foot4}. 
Therefore $\gamma \gamma \to \sneuipm$ in a future PLC,
remains as the sole 
process for production of a sneutrino as
the only supersymmetric particle within $\rp$ MSSM.

\end{itemize}

Within the above scenarios, for $\tilde\nu \to \gamma \gamma~,~gg$,  
we can assume without loss of generality, that 
only flavor diagonal $\tilde\nu$ couplings 
are present. 
Furthermore, for definiteness, in both scenarios we will only consider 
couplings 
of the $\mu$--sneutrino, {\it i.e.} $i=2$, although our results 
hold for the $e$--sneutrino as well; for $i=3$ 
the results for $\sneut$ decay to 
$\gamma \gamma$ will have only quarks 
in the loop, as
the $\tilde\nu^\tau \tau \tau$ 
coupling $\lambda_{333}$ is forbidden.

Before presenting the results of our study 
we remark that, just as in the case of 
Higgs reactions, higher  order corrections 
(mainly QCD), 
may be substantial \cite{hepph9705337}
for both decay widths and production cross--sections. 
Since such corrections have
not been calculated for sneutrinos, and since the discussion here is
exploratory, all higher order corrections will be ignored.  
The values of the parameters used here are: 
$m_\tau=1.8$ GeV, $m_b=4.5$ GeV, $\alpha=1/128$ and
$\alpha_s=0.118$. 

We note that bounds on the sneutrino masses can be obtained 
without reference to a
specific $\rp$ scenario using sneutrino pair production at, for example,
LEP2. This can be done, for instance,  
through $R_P$--conserving MSSM interactions (see {\it e.g.}
\cite{plb364p27}), and the subsequent decays into four fermion
states, {\it i.e.} $\tilde \nu \tilde \nu \to \tau  \tau  \tau  \tau,~
bbbb,~bb \tau  \tau$ 
in scenario 1, or $\tilde \nu \tilde \nu \to \tau  \tau  \tau  \tau$    
in scenario 2. However, the pair production cross--section 
strongly depends on the values of the $R_P$--conserving MSSM parameters. 
Thus, one cannot exclude light sneutrinos with masses $\gsim 50$ GeV
from current LEP2 data (see {\it e.g.}, \cite{hepex9712013}).
Recently resonant sneutrino production has
been searched for in \cite{hepex9808023}; 
note however that non of their scenarios
is the same as ours.

There are also bounds for the $\rp$ couplings
relevant to the above scenarios for $i=2$,
{\it i.e.} on $\lambda^{\prime}_{233}$ and $\lambda_{233}$. 
These bounds are usually given at the $1\sigma$  or $2\sigma$
level and are deduced by using some simplifying assumptions, {\it e.g.,} 
only one coupling at a time is assumed 
to be non-zero (see \cite{rpreview} and references therein).
Furthermore, the bounds are usually presented for $m_{\tilde f}=100$ GeV,
where $\tilde f \neq \tilde\nu$ is the sfermion
involved in the process employed to obtain the bounds, and such
constraints become weaker as  $m_{\tilde f}$ increases.
For a notable exception see \cite{hepex9808023}.

The $1\sigma$ upper limit on $\lambda^{\prime}_{233}$ 
is about $0.4$ for 
$m_{\tilde b} =100$ GeV; it is derived (see 
\cite{mpla10p1583} and its update in \cite{rpreview}),
from the data for the ratio $\Gamma(Z \to {\rm hadrons})/
\Gamma(Z \to \mu^+ \mu^-)$. The upper limit rises (practically linearly)
with $m_{\tilde b}$, reaching  ${\cal O}(1)$ 
for $m_{\tilde b}$ around $500$ GeV \cite{mpla10p1583}.
We can therefore take $0.5< \lambda^{\prime}_{233} < 1.5$, 
without violating any existing bound; this will be the
range investigated within scenario 1.

The $1\sigma$ upper limit 
on $\lambda_{233}$ was extracted (see \cite{resold} 
and its update in \cite{rpreview})
from the ratio 
$R_\tau \equiv \Gamma(\tau \to e \nu \bar\nu)/
\Gamma(\tau \to \mu \nu \bar\nu)$ and it scales with the stau mass as 
$\lambda_{233}<0.06 (m_{\tilde\tau}/100~{\rm GeV})$.  
Both $\lambda_{133}$ and $\lambda_{233}$ contribute to $R_\tau$ 
where their contributions appear with a relative minus sign 
\cite{resold}. 
Therefore, either by assuming that $m_{\tilde\tau} \gsim 500$ GeV and 
requiring an effect larger than 1$\sigma$, or assuming that there is  
a (possibly partial)
cancellation between the contributions of $\lambda_{133}$ 
and $\lambda_{233}$ to $R_\tau$, $\lambda_{233}=1$ is not ruled
out.  Hereafter, we fix the value of  $\lambda_{233}$ to unity,  
for both scenarios.

We now consider the prospects of discovering sneutrinos at the LHC 
via their decay to a pair of photons 
within scenario 1, then briefly 
comment on the corresponding effects at the Tevatron.
Within the present scenario,  the $\tilde\nu bb$ 
coupling is much larger than the   
$hbb$ one, and the effective $\tilde\nu gg$ coupling
is smaller than the $hgg$ one. Therefore, at the LHC, single sneutrinos
are expected to be produced mainly from the parton processes
listed below, which result from the $\tilde\nu bb$ coupling,
while the Higgs is considered to be
dominantly produced through the $gg$ fusion. 
The leading processes contributing
to the inclusive single sneutrino production, $pp \to \sneupm +X$ at
the LHC, are:

\begin{enumerate}
 
\item $s$--channel resonant sneutrino production: $b \bar b \to \sneupm$.

\item Associated production of sneutrino and a $b$--jet:
$b g \to b \sneupm$, where the $\tilde\nu$ is obviously 
either emitted from the outgoing $b$
(an $s$--channel process), or from the incoming $b$ 
(a $t$--channel process). In both cases one needs to 
add the corresponding cross--sections with $b \longrightarrow \bar b$,
which is equivalent to multiplying the $b$--quark result by 2.
This is the analog to the process 
$e \gamma \to \tilde\nu e$,
discussed in \cite{plb420p307}.

\item Associated production of sneutrino and a $g$--jet:
$b \bar b \to \sneupm g$, again where the sneutrino is 
emitted either from the $b$ or from $\bar b$.

\end{enumerate}

We have also studied the $2 \to 3$ subprocess
 $gg \to \sneupm b \bar b$. Naively, this process is a source for
 large logarithms. However, to avoid double counting, since the logs
 are already included in the definition of the b-quark parton distributions
 \cite{doublecount}, we have done a rough estimate of the rates
 for the $2 \to 3$ subprocess without including these logs.
 The remaining contributions for the $2 \to 3$ subprocess is estimated to be
much smaller than the $2 \to 1$ and $2 \to 2$ processes 
mentioned above and therefore is not being included in our calculations.

The cross--sections for $\sneumpm$ production in a hadron collider 
are then obtained
\cite{colliderphys} by folding the parton--level
cross--sections  with the relevant
parton distribution functions in the beams,
neglecting all higher order 
corrections, as mentioned above. 
We follow this procedure, employing
the CTEQ4M parameterization \cite{cteq} and
find that the cross-sections for the first and second processes
above are approximately equal to each
other and larger than 
the third process by about an
order of magnitude; nevertheless, for completeness, the latter is 
also included \cite{foot5}.

Since sneutrino and Higgs production rates and decays are
expected to have similar higher order corrections
and are expected to be subjected to 
comparable experimental cuts, 
and since the expected statistical significance of the $h \to \gamma \gamma$ 
signal at the LHC is known \cite{egede}, the ratio

\be
R=\frac{\displaystyle{\sum_{s=+,-}}\sigma (pp \to \sneums\ + X){\rm Br}
(\sneums \to \gamma \gamma)} 
{\sigma (pp \to h + X) {\rm Br}
(h \to \gamma \gamma)} \label{ratior}~,
\ee

\n will provide a simple guide for the possibility of using
$\tilde\nu \to \gamma \gamma$ as a detection channel for 
sneutrinos at the LHC. The plot of $R$ as a function of $m_{\tilde\nu}=m_h$ for
$\sqrt s=14$ TeV (corresponding to the LHC) is presented in Fig. 1 
where, as throughout this paper,
we take  $m_{\tilde\nu} \equiv m_{\sneump} = m_{\sneumm}$.
We note that the branching  ratios $\Gamma(\sneump \to \gamma \gamma) 
\approx \Gamma(\sneumm \to \gamma \gamma)$ 
within $\sim 10\%$ and the production cross--sections for
$\sneumpm$ are equal up to $\sim 50\%$. Cross--sections and 
branching ratios were calculated to lowest order in EW
 and QCD, as mentioned before, and without cuts.
Results for three values of $\lambda^{\prime} \equiv   
\lambda^{\prime}_{233}$, all with $\lambda \equiv \lambda_{233}=1$
are displayed; note that we assume that both couplings appear
with the same sign.

\begin{figure}[htb]
\psfull
 \begin{center}
  \leavevmode
  \epsfig{file=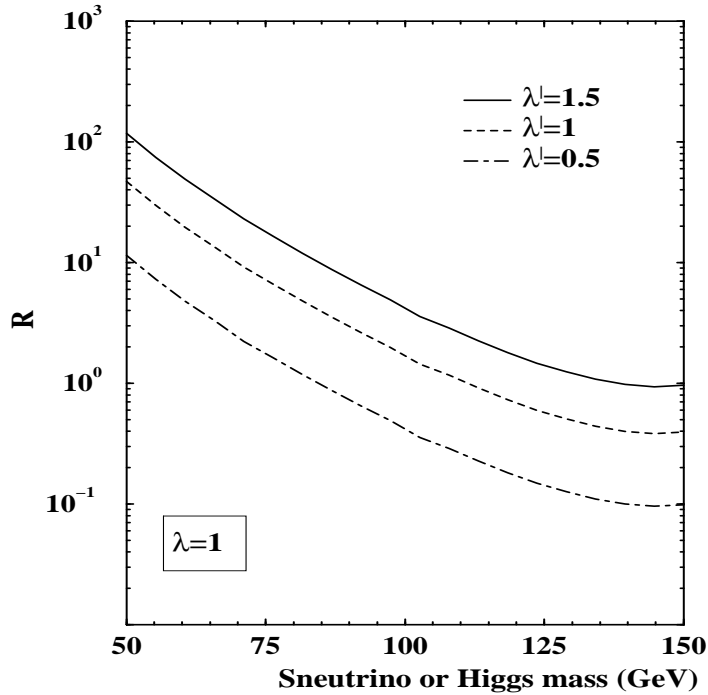,height=9.5cm,width=10cm,bbllx=0cm,bblly=2cm,bburx=20cm,bbury=25cm,angle=0}
 \end{center}
\caption{\emph{The ratio $R$ of the production cross--
sections $\times$ branching ratios for decays to $\gamma \gamma$,
between sneutrinos and Higgs ($R$ is defined in Eq. \ref{ratior}), 
at the LHC with $\sqrt s=14$ TeV,
as a function of the mass $m_h=
m_{\sneump}=m_{\sneumm}$, in scenario 1 (see text) with $\lambda=1$ 
and $\lambda^{\prime}=0.5$ (dash-dot), 1 (dashed) or 1.5 (solid). 
Only leading order terms 
are kept in $R$, also no cuts are imposed.}}
\label{fig1}
\end{figure}

We approximate the branching ratio of $\sneumpm \to \gamma \gamma$ by:

\be
{\rm Br}(\sneumpm \to \gamma \gamma)= 
\frac{\Gamma(\sneumpm \to \gamma \gamma)}{\Gamma(\sneumpm \to b \bar b) + 
\Gamma(\sneumpm \to \tau^+ \tau^-)+
\Gamma(\sneumpm \to \tilde\chi^+ \ell) + 
\Gamma(\sneumpm \to \tilde\chi^0 \nu)} \label{br}~,
\ee

\n where (see Barger {\it et al.} in \cite{resold}):  

\bea
&&\Gamma(\sneumpm \to \tilde{\chi}^+ \ell) 
\sim {\cal O} \left[ 10^{-2} m_{\sneumpm} \times 
\left(1 - m_{\tilde {\chi}^+}^2/m_{\sneumpm}^2 \right)^2 \right]
\label{sneutochiplus}~,\\
&&\Gamma(\sneumpm \to \tilde{\chi}^0 \nu) 
\sim {\cal O} \left[ 10^{-2} m_{\sneumpm} \times 
\left(1-m_{\tilde {\chi}^0}^2/m_{\sneumpm}^2\right)^2 \right] 
\label{sneutochizero}~.
\eea 

\n Evidently, with $\lambda^{\prime}=1$, for example, the $R_P$--conserving 
decay channels of the sneutrino, if open, are always   
smaller than the $\rp$ decays to a pair of $b$--quarks: 
 
\be
\Gamma(\sneumpm \to b \bar b)= 
\left( \lambda^{\prime}_{233} \right)^2 \frac{3}{16 \pi} 
m_{\sneumpm} \label{sneutodd}~.
\ee

\n Therefore, for simplicity, we take (conservatively),   
$\Gamma(\sneumpm \to \tilde\chi^+ \ell) + 
\Gamma(\sneumpm \to \tilde\chi^0 \nu) = 10^{-2} m_{\sneumpm}$~, ignoring 
the phase--space factors in Eqs.~\ref{sneutochiplus} and 
\ref{sneutochizero}.  

As can be seen from Fig. 1, the ratio $R$ 
in Eq.~\ref{ratior}, obeys $R \gsim 1$ for  
$m_{\tilde\nu} \lsim 85,110,140$ GeV, when $\lambda^{\prime}
\gsim 0.5,1,1.5$, respectively. Moreover, it is interesting to note 
that $R \gsim 10$ with $\lambda^{\prime}
\gsim 0.5,1,1.5$ for $m_{\tilde\nu} \lsim 50,70,85$ GeV, respectively.
Now, for $pp \to h +X$ followed by $h \to \gamma \gamma$ 
the values for $S/\sqrt B$ range between 2.3 and 7.1 for $m_h$
between 80 and 140 GeV \cite{egede},
where $S$ is the signal for single Higgs production and its
subsequent decay into two photons at the LHC with higher order
corrections included, and $B$
is the QCD background. Since the higher order
corrections are expected to be similar for the numerator and denominator
in $R$, therefore the results in Fig. 1 are encouraging
and a further   
study of the two photon decay modes of sneutrinos produced
singly in  $\rp$ MSSM is warranted. Later we will
consider the QCD background to $pp \to \sneumpm + X
\to \gamma \gamma + X$. The conclusions are similar to the ones above, 
though they should be 
verified by including radiative corrections.

In Fig. 2a we plot ${\rm Br}(\sneump \to \gamma \gamma)$, 
which is approximately equal to
 ${\rm Br}(\sneumm \to \gamma \gamma)$,
as a function of the sneutrino mass, in scenario 1.
It is interesting to note that
while ${\rm Br}(h \to \gamma \gamma)$ sharply falls 
once $m_h \gsim 2m_W$, from $\approx 10^{-3}$ around $150$ GeV, to
$\approx 10^{-7}$ at $550$ GeV,
${\rm Br}(\sneumpm \to \gamma \gamma)$ smoothly 
drops only by about an order of magnitude as one goes from 
$m_{\tilde\nu} =50$ to $550$ GeV, where it is $\approx 10^{-7}$.

In Fig. 2b we show, again for scenario 1,
the total number of $\mu$--sneutrinos, both
$\sneump$ and $\sneumm$, produced at the LHC with
a high luminosity of 100 fb$^{-1}$.
Taking $\lambda^{\prime}=1$, and comparing
the expected number of sneutrinos
produced singly at the LHC (mainly through the $b \bar b$ and $bg$
fusion mechanisms),
with the expected 
number of Higgs produced 
(predominantly through $gg \to h$ \cite{hepph9803257}),
we find that for $m_{\tilde\nu}=m_h=100$ GeV the number of 
sneutrinos is more than two orders of magnitude larger, while  
for $500$ GeV it is about an order of magnitude larger.

\begin{figure}[htb]
\psfull
 \begin{center}
  \leavevmode
  \epsfig{file=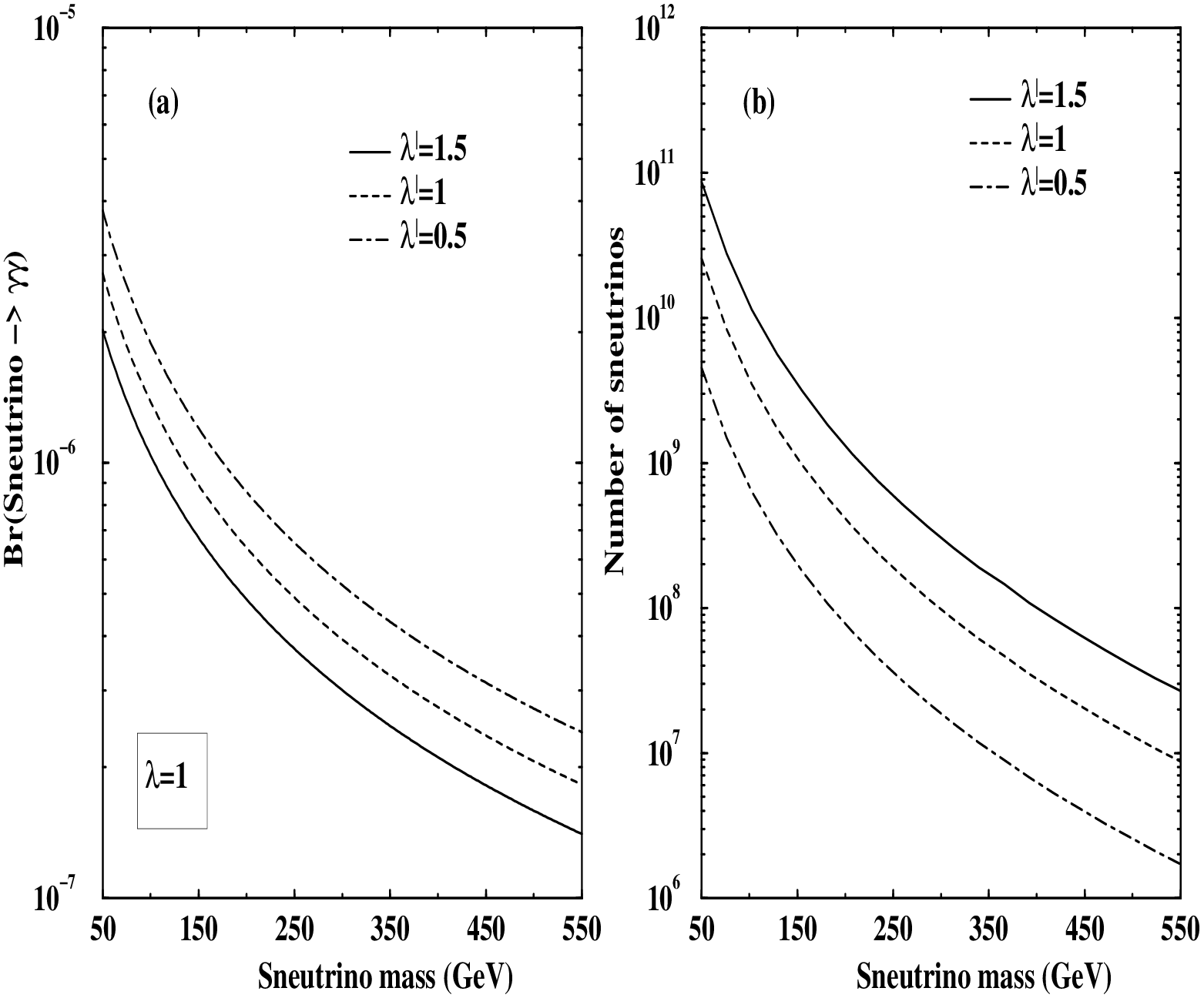,height=8.5cm,width=13cm,bbllx=0cm,bblly=2cm,bburx=20cm,bbury=25cm,angle=0}
 \end{center}
\caption{\emph{(a) The branching ratio 
${\rm Br}(\sneump \to \gamma \gamma)$; the numbers for $\sneumm$ 
are very similar.
(b) The number of $\mu$--sneutrinos 
with CP~$=+$, plus the number of $\mu$--sneutrinos 
with CP~$=-$,
at the LHC with $L=100$ fb$^{-1}$.
Both figures are 
in scenario 1 (see text). See also caption to Fig.~1.}}
\label{fig2}
\end{figure}

In order to better estimate the feasibility 
of detecting the sneutrino through its decay to 
a pair of photons one has to study the signal to background ratio, 
{\it i.e.}  $pp \to \sneumpm +X \to \gamma \gamma + X$ versus 
$pp \to \gamma \gamma + X$ from the continuum.
At the LHC, as a result of
the high $gg$ luminosity, the box graph contribution to                       
$g g \to \gamma \gamma$ is comparable to the tree-level 
$q \bar q \to \gamma \gamma$ one
and also has to be considered.  
Comprehensive background analysis is beyond the 
scope of this work 
(this can be found, for example, in \cite{egede} 
for $h \to \gamma \gamma$ at the LHC). For the purposes
of this paper it suffices to calculate
$d\sigma/dM_{\gamma \gamma} (q \bar q \to \gamma \gamma)$, 
where $M_{\gamma \gamma}$ is the invariant mass of the photon pair, 
and multiply it by a factor of two to account for the 
box mediated subprocess $g g \to \gamma \gamma$ \cite{jose}. 
The number of background $\gamma \gamma$ events is therefore taken here 
as:

\be 
B= 2 \times \frac{d\sigma}{dM_{\gamma \gamma}} 
(q \bar q \to \gamma \gamma) \times 
\Delta M_{\gamma \gamma} \label{bb}~,
\ee

\n where 
$\Delta M_{\gamma \gamma}$ is the mass resolution bin for the
 reconstruction of the 
$\gamma \gamma$ invariant mass which we take to be 
$\Delta M_{\gamma \gamma} = 10^{-2} M_{\gamma \gamma} $, {\it i.e.} 
$1\%$ accuracy in measuring $M_{\gamma \gamma} $ is assumed.

The signal for $m_{\sneump} \approx m_{\sneumm}$, 
is given by:

\be
S= \left[ \sum_{s=+,-} \sigma(pp \to {\sneum}_s +X) \times L \times 
{\rm Br}({\sneum}_s \to \gamma \gamma) \right] \times (1-t) \label{ss} ~.
\ee

\n We have included the factor $(1-t)$ in Eq.~\ref{ss} to take into account 
the reduction in signal within one bin since the sneutrino width is larger 
than $1\%$ of its mass.
Specifically, we choose $t=1/2$ thus decreasing the signal by half. 

In Fig. 3 we show the statistical significance $S/\sqrt B$ 
for the process $pp \to \sneumpm +X \to \gamma \gamma +X$ 
as a function of the sneutrino mass.
In calculating both $S$ and $B$ we employ a cut on the photon scattering 
angle $|\cos\theta|<0.5$ and 
again we take a high yearly luminosity at the LHC ($L=100$ fb$^{-1}$).
We find that, with $\lambda^{\prime}=0.5$, $S/\sqrt B > 1$ 
only if $m_{\sneutpm} \lsim 70$ GeV, thus
in this case the outlook is not that 
optimistic.
We therefore discuss the
numerical results only for $\lambda^{\prime}=1,1.5$.  
We observe from Fig. 3 that, for $\lambda^{\prime}=1$,  
a $1\sigma$ signal is possible at the LHC for 
$m_{\sneumpm} \lsim 125$ GeV. If $\lambda^{\prime}=1.5$, then 
a $1\sigma$ sneutrino signal
through $\sneumpm \to \gamma \gamma$ may be 
possible for $m_{\sneumpm} \lsim 180$ GeV.
Furthermore, the $3\sigma$ discovery sensitivity seems 
attainable for 
$m_{\sneumpm} \lsim 85,120$ GeV when $\lambda^{\prime}=1,1.5$,
respectively. It is interesting to note
that the discovery ranges for both the SM Higgs and
the sneutrino through their $\gamma \gamma$ decay
modes at the LHC, are about the same.

\begin{figure}[htb]
\psfull
 \begin{center}
  \leavevmode
  \epsfig{file=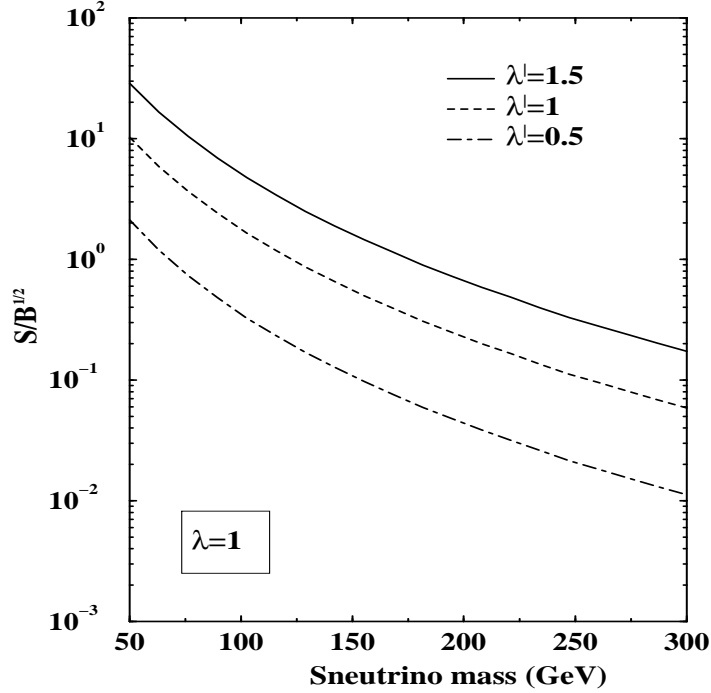,height=9.5cm,width=10cm,bbllx=0cm,bblly=2cm,bburx=20cm,bbury=25cm,angle=0}
 \end{center}
\caption{\emph{The statistical significance $S/\sqrt B$,   
for a signal from $pp \to \sneum +X \to \gamma \gamma +X$, 
as a function of the $\sneum$ mass at the LHC with $L=100$ fb$^{-1}$,
in scenario 1 (see text).
A cut on the photon scattering 
angle $|\cos\theta|<0.5$ is imposed. The signal is defined in  Eq.~\ref{ss} 
and the background is given in Eq.~\ref{bb}.
See also caption to Fig.~1.}}
\label{fig3}
\end{figure}

We close this discussion of single sneutrino production in hadron 
colliders within scenario 1, with a comment about single
sneutrino production at the Tevatron. As mentioned before, due to
the small probability of finding a $b$--quark or gluon
in a proton (or anti-proton) at $\sqrt s=2$ TeV, the $b \bar b$ 
and $bg$ luminosities 
at the Tevatron,
are too small to allow detection of sneutrinos through their
two photons decay mode. We can, of course, envisage
a modified scenario in which the sneutrino,  which  for definiteness
is again taken
as $\sneum$, is produced from the valence $d {\bar d}$ annihilation.
However, it is unlikely that $\lambda^{\prime}_{211}$ 
is of ${\cal O}(1)$, 
as required in order to render the production rate
large enough. This is due to a gauge hierarchy argument,
and to the fact that, in this case, there is no possibility of a
cancellation between the contributions of two 
$\lambda^{\prime}$ couplings \cite{resold}.

Finally, let us investigate scenario 2 and its implications for 
$e$--sneutrino and $\mu$--sneutrino 
production in a future PLC. As mentioned earlier, 
if only $\lambda_{i33} \neq 0$, for $i=1,2$ 
then the only production 
mechanism of an $s$--channel sneutrino is via $\gamma \gamma$ fusion, 
$\gamma \gamma \to \sneuipm$.
Following \cite{prd48p4018}, we find that the number of 
$\sneuipm$ produced in a polarized PLC is (see also \cite{hepph9206262}):
\bea
N_{\sneuipm}&=&\left.\frac{d L_{\gamma \gamma}}{dW_{\gamma \gamma}} 
\right|_{W_{\gamma \gamma} =m_{\sneuipm}} \times 
\frac{4 \pi^2 \Gamma(\sneuipm \to \gamma \gamma)}{m_{\sneuipm}^2} 
\times (1 + h_1 h_2) \simeq 1.54 \times 10^4 \times \cr
&&\left(\frac{ L_{ee}}{{\rm fb}^{-1}} \right) 
\left(\frac{E_{ee}}{{\rm TeV}}\right)^{-1} \left( \frac{\Gamma(\sneuipm 
\to \gamma \gamma)}{{\rm keV}}\right) \left(\frac{m_{\sneuipm}}{{\rm GeV}} 
\right)^{-2} F(m_{\sneuipm}) \times (1 + h_1 h_2) \label{nplc}~,
\eea
   
\n where $h_1,h_2$ are the helicities of the initial photons and
 $E_{ee}$ and $L_{ee}$ are the $e^+e^-$ machine energy and 
yearly integrated luminosity, respectively. 
$F(W_{\gamma \gamma}) = (E_{ee}/L_{ee}) 
dL_{\gamma \gamma}/dW_{\gamma \gamma}$ 
depends on the machine parameters and is of ${\cal O}(1)$ \cite{prd48p4018}; 
for simplicity we take $F(W)=1$, \cite{hepph9206262}.
Note again that for $m_{\sneuip} = m_{\sneuim}$ we have
$\Gamma(\sneuip \to \gamma \gamma) \simeq 
\Gamma(\sneuim \to \gamma \gamma)$, so that
the number of $(\sneuip +\sneuim)$ produced 
is $N_{(\sneuip +\sneuim)} \simeq 2 N_{\sneuipm}$.
In what follows we consider $\sneumpm$ production 
with the relevant coupling $\lambda_{233} \neq 0$;
the analysis for $\tilde\nu^e_{\pm}$ 
production is similar. 

In Table 1 we give, for $m_{\sneump} = m_{\sneumm} = 50 - 300$ GeV, 
the scaled partial width 
$\Gamma(\sneumpm \to \gamma \gamma)/ \lambda_{233}^2$ and
the expected number of $\sneump +\sneumm$ also scaled by 
$\lambda_{233}^2$. (Recall that 
$\Gamma(\sneumpm \to \gamma \gamma) \propto \lambda_{233}^2$; 
see Eq.~\ref{gammagamma} and assume scenario 2 with $i=2$.) 
In the table, 
$N_{(\sneump +\sneumm)}/ \lambda_{233}^2$ is given for a polarized PLC with 
initial photon helicities \cite{foot6} 
 $h_1h_2=1$ and for two $e^+ e^-$ luminosity values 
of $L_{ee}= 20$ fb$^{-1}$ and a high luminosity PLC with 
$L_{ee}= 100$ fb$^{-1}$, both for $E_{ee}=0.5$ TeV.  
Evidently, for $\lambda_{233} =1$, from
thousands to hundreds of sneutrinos 
with masses 50 -- 300 GeV, respectively, may be produced in a PLC 
with an integrated luminosity of $L_{ee} =100$ fb$^{-1}$. 
Similarly, hundreds to tens of sneutrinos may be produced if  
$L_{ee} =20$ fb$^{-1}$ within the same $\sneutpm$ mass range.  
\begin{table}
\begin{center}
\caption[first entry]{\emph{
The $\sneump \to \gamma \gamma$ width 
in keV (the widths for $\sneumm \to \gamma \gamma$
are larger by $\sim 10\%$)
and the approximate 
number of $\sneumpm$ produced in a
future $\gamma \gamma$ collider ,  
scaled by $\lambda_{233}^2$, for $E_{ee}=0.5$ TeV. 
We take $L_{ee}=20$ fb$^{-1}$ and $L_{ee}=100$ fb$^{-1}$, with
$m_{\sneumpm}=50 - 300$ GeV. All entries are within scenario
2 (see text).}

\bigskip

\protect\label{table1}}
\begin{tabular}{|r||r|r|r|r|r|r|} \cline{1-7}
$m_{\sneumpm}$, GeV $\Longrightarrow$& 50 & 100& 150 & 200 & 250 & 300 \\ \hline \hline 
$\Gamma(\sneump \to \gamma \gamma)/\lambda_{233}^2$, keV&2.8&2.7 & 2.5 & 2.4&2.2&2.1 \\ \hline 
$N_{(\sneump + \sneumm)}/\lambda_{233}^2$ ($L_{ee}=20$ fb$^{-1}$) & 2700& 650& 270 & 150 & 90 & 60 \\ \hline 
$N_{(\sneump + \sneumm)}/\lambda_{233}^2$ ($L_{ee}=100$ fb$^{-1}$) & 14000& 3300& 1400 & 730 & 440 & 290 \\ \hline 
\end{tabular}
\end{center}
\end{table}

Given the event rates in Table 1, one can estimate the statistical 
significance 
of the sneutrino signal in a PLC. Within scenario 2 with
$\lambda_{233} \neq 0$, the $\mu$--sneutrino will decay predominantly to
$\tau^+ \tau^-$, 
if we assume, for simplicity, that its $R_P$--conserving decays
are either suppressed by 
phase--space factors in (\ref{sneutochiplus}) and (\ref{sneutochizero}),
or are kinematically inaccessible.
With these assumptions, 
${\rm Br}(\sneumpm \to \tau^+ \tau^-) \sim 1$ with effectively 
no dependence on $\lambda_{233}$.
The main background to $\gamma \gamma \to \sneumpm \to \tau^+ \tau^-$ 
is therefore from the continuum $t$--channel tree--level process 
$\gamma \gamma \to \tau^+ \tau^-$. However, a significant reduction 
of this background is achieved
if the $\tau^+ \tau^-$ pair are restricted to be in a $J_z=0$ state by using 
polarized photon beams \cite{prd48p4018,hepph9206262}. 

In \cite{prd48p4018} the number of tree--level $t$--channel exchange
continuum  $\gamma \gamma \to c \bar c,~b \bar b$ 
events, as a function of the $q \bar q$ ($q=c,~b$) 
invariant mass and in 10 GeV mass bins, was calculated in order to estimate 
the background to $\gamma \gamma \to h \to c \bar c,~b \bar b$. 
A cut on the scattering angle $|\cos\theta|<0.7$ was imposed, and  
only $q \bar q$, 
$J_z=0$ states were taken into account.
We will use these results 
to estimate our $\gamma \gamma \to \tau^+ \tau^-$ background. Note 
that since 
$\Gamma_{\sneumpm} < 10$ GeV, we will assume that
all the sneutrino events fall in one bin.

To a good approximation,
$\sigma(\gamma \gamma \to \tau^+ \tau^-)$ can be calculated from 
$\sigma(\gamma \gamma \to c \bar c)$ by multiplying the latter by 
$(Q_c^4 N_c)^{-1}= 27/16$, and disregarding the very mild
change due to the replacement $m_c \to m_{\tau}$ \cite{foot7}.
The significance of the $\sneumpm$ signal is given by $S/\sqrt B$, 
where $S=N_{(\sneump +\sneumm)}\mid_{|\cos\theta|<0.7}
\times {\rm Br}(\sneumpm \to \tau^+ \tau^-)$ 
and $B=N(\gamma \gamma \to \tau^+ \tau^-)\mid_{|\cos\theta|<0.7}$ from the
tree--level process. Using the results 
in \cite{prd48p4018} we find that
for $L_{ee}=20$ fb$^{-1}$ and 
$m_{\sneumpm}=50,~100,~150,~200$ GeV, $S/\sqrt B \simeq 46,~8,~3,~4$, 
respectively. 
For $L_{ee}=100$ fb$^{-1}$ the corresponding numbers are, 
$S/\sqrt B \simeq 103,~17,~7,~9$, respectively.
We also note that with $M_{\gamma \gamma} \gsim 200$ GeV the background 
event rates sharply drop such that there are fewer than 
$ \sim 85$ background events for $M_{\gamma \gamma} \gsim 300$ GeV.
 Therefore, even with a heavy sneutrino of mass $\sim 300$ GeV, 
$S/\sqrt B \gsim 4$ for $L_{ee}=20$ fb$^{-1}$ and $S/\sqrt B \gsim 10$ for 
$L_{ee}=100$ fb$^{-1}$. This can be compared
with the $s$--channel neutral Higgs case, 
where the statistical significance of the signal from 
$\gamma \gamma \to h \to c \bar c,~b \bar b$ drops below $\sim 3$ for 
$m_h \gsim 160$ GeV (when $L_{ee}=20$ fb$^{-1}$) \cite{prd48p4018} 
due to the opening of the decay channel 
$h \to VV$, $V=W$ or $Z$, and 
even before that to $V V^*$. For a heavier Higgs, 
$h \to t \bar t$ 
becomes important. 
Higgs bosons can then be detected in a PLC through
these new decay modes \cite{hepph9801359}, which have their own
backgrounds and are not available for  
sneutrino decays at tree--level (having neglected mass mixings in the
superpotential). Therefore, both $\tilde\nu$ 
and $h$ may be resonantly produced at a future photon--photon
collider, then observed through their decays.
 
To summarize, we have demonstrated that 
within certain scenarios
in $\rp$ MSSM, the loop--induced
decay $\tilde\nu \to \gamma \gamma$ may be used as a
tool for detecting singly produced
sneutrinos at a high luminosity LHC 
over a significant $\tilde\nu$ mass range,
if the relevant $\rp$ couplings are large enough, yet still within
their experimentally allowed bounds. In addition we have discussed, 
at some length,
resonant sneutrino production in a $\gamma \gamma$ collider.

At the LHC the main single sneutrino 
production mechanisms would be $b \bar b$ and $bg$
fusion, while the Higgs will be produced via $gg$ fusion.
At the Tevatron, sneutrino production 
through $d \bar d$ fusion (irrelevant for resonant 
Higgs production) and its decay, via the two
photon mode, appears too small. 
Resonant sneutrino production in a hadron collider through $q \bar q$ fusion,
has already been discussed in the literature 
\cite{rpreview,resold,sneuresonant,ourpapers}. Here we 
add two processes, where one of them, namely $bg \to \tilde\nu b$ 
is as significant as $b \bar b \to \tilde\nu$
at the LHC, and suggest $\tilde\nu \to \gamma \gamma$ as 
a relatively clean decay mode of sneutrinos as a signal for their
detection.
Though sneutrinos may be more abundantly produced than Higgs bosons
in hadron colliders, their branching ratio to $\gamma \gamma$
are usually smaller (except for very high masses). These 
two effects thus
compensate each other. 
The relatively clean 
$\gamma \gamma$ mode remains a promising prospect for 
detection, at least for masses $\lsim 125-180$ GeV for 
$\lambda^{\prime}=1-1.5$, as can be seen from Figs. 1 and 3. 

Sneutrinos can also be
produced in future $\gamma \gamma$ colliders with a statistically
significant signal, over a wide $m_{\tilde\nu}$ range. 
In particular, we have investigated a scenario in which
all the couplings of the $\lambda^{\prime}$ type vanish 
and only $\lambda_{i33} \neq 0$. In this case,  $\gamma \gamma$
colliders will be the only venue to produce resonant $s$--channel
sneutrinos. In fact, the effect of a new $\rp$ one-loop 
$\tilde\nu \gamma \gamma$ coupling is much more pronounced in a 
$\gamma \gamma$ collider than at the LHC. The reason is that 
$\tilde\nu$ production via $\gamma \gamma$ fusion is proportional to 
$\Gamma(\tilde\nu \to \gamma \gamma)$ which is only about one order 
of magnitude smaller then $\Gamma(h \to \gamma \gamma)$, whereas 
$pp \to \tilde\nu +X\to \gamma \gamma +X$ is proportional to 
Br$(\tilde\nu \to \gamma \gamma)$ which is about three orders of magnitude 
smaller then Br$(h \to \gamma \gamma)$ in the interesting mass range,  
$m_{\tilde\nu}~{\rm or}~m_h \lsim 140$ GeV.    

The situation is less optimistic with
regards to uses of $\tilde\nu \to gg$: for sneutrino detection
it will be (as in the Higgs case) swamped by the QCD background,
while as far as production at the LHC goes, $gg \to \tilde\nu$ will 
be overshadowed by $b \bar b \to \tilde \nu$ and $bg \to \tilde\nu +b$, 
unlike the Higgs case where 
it will be mainly produced through $gg$ fusion.  
 
Loop--induced sneutrino decays to $WW$ and $ZZ$, 
are expected at the same order as the decays to $\gamma \gamma$
(disregarding $L_i H_u$ terms).
These decays may also 
be useful for $\tilde\nu$ production
({\it e.g.} through $WW$ fusion), or detection ({\it e.g.} in
$\tilde\nu \to ZZ$). 

A few directions for future research are listed below:

\begin{enumerate}

\item EW and QCD corrections to the lowest order processes presented
here may need to be calculated.

\item More realistic background estimates have to be performed,
including signal to background ratios for the competing
processes $pp \to \tilde\nu + X \to b \bar b\,,~ \tau^+ \tau^- + X$. 
Similarly, the 
interesting triple--fermionic final states in 
$pp \to \tilde\nu + X \to b b \bar b\,,~ b \tau^+ \tau^- + X$ in which 
the single sneutrino from $bg \to \tilde\nu +b$ decays to 
$b \bar b$ or $\tau^+ \tau^-$, should be studied.

\item Can one gain much by requiring 
that a high $p_t$ $b$--jet
accompany the two photons, thus enhancing the sensitivity
to the processes $b g \to \tilde\nu +b$? In this case, an extra trigger, 
{\it i.e.,} the distinctive high $p_t$ $b$--jet, is present. 

\item The importance of the loop processes $\tilde\nu \to 
WW,~ZZ$ remains  to be assessed.
\end{enumerate} 

We acknowledge partial support from the US Israel BSF (G.E. and A.S.) 
and from the US DOE contract numbers DE-AC02-76CH00016(BNL), 
DE-FG03-94ER40837(UCR). G.E. thanks the Israel Science Foundation and the 
VPR Fund at the Technion for partial support 
and members of the HEP group at UCR for their hospitality.
    

%

%
%
%
%






\begin{thebibliography}{99}
\bibitem{foot2} $F_{1/2}^+$ is the function commonly denoted by $F_{1/2}$.
\bibitem{foot3} Such a collider is contemplated as an adjunct to an $e^+ e^-$
next linear collider, where high energy photons will be produced by
Compton backscattered laser light off the initial beams \cite{photonphoton}.
\bibitem{foot4} We note that 
if all $\lambda^{\prime}_{ijk}=0$, then $\Gamma(\sneui \to gg) = 0$ 
at one loop too.
\bibitem{foot5} Sneutrino production via $gg \to \sneumpm$, 
although negligible compared to the other processes,
is included as well.
\bibitem{foot7} This is  
exemplified by the fact that 
the ratio between $b \bar b$ and $c \bar c$ is approximately
given by the ratio of the 4th power of the charges
\cite{prd48p4018}.
\end{thebibliography}
\end{document}